\newcommand{\beq}{\begin{equation}}
\newcommand{\eeq}{\end{equation}}

\def \xray {\hbox{X--ray} }

\def \logTd6 {\hbox{log$( T/6 \kev)$} }

\def\myputfigure#1#2#3#4#5%
{\vskip#5pt\makebox[0pt]{\hskip#2in
\includegraphics[width=#3\textwidth]{#1}}\vskip#4pt\hfill}

\def \hmpc       {{$h^{-1}$\rm\ Mpc}}

\def \kms       {\hbox{ km s$^{-1}$}}

\def \kev       {{\rm\ keV}}

\def \hmsol     {h^{-1}{\rm\ M}_\odot}
\def \hMpc      {h^{-1}{\rm\ Mpc}}
\def \hkpc      {h^{-1}{\rm\ kpc}}
\def \mkms      {\rm\ km\ s^{-1}}

\documentstyle[emulateapj,onecolfloat,natbib]{article}
\input epsf

\begin{document}
\twocolumn[
\title{Cold Fronts in CDM Clusters}
\author{Daisuke Nagai and Andrey V. Kravtsov}
\affil{Center for Cosmological Physics,
University of Chicago, Chicago  IL 60637}
\affil{Department of Astronomy and Astrophysics,
University of Chicago, 5640 S Ellis Ave, Chicago  IL 60637}

\begin{abstract}
  Recently, high-resolution {\sl Chandra} observations revealed the
  existence of very sharp features in the X-ray surface brightness and
  temperature maps of several clusters \citep{vikhlinin01}.  These
  features, called {\em ``cold fronts''}, are characterized by an
  increase in surface brightness by a factor $\gtrsim 2$ over
  10-50~kpc, accompanied by a {\it drop} in temperature of a similar
  magnitude. The existence of such sharp gradients can be used to put
  interesting constraints on the physics of the intracluster medium
  (ICM), if their mechanism and longevity are well understood. Here,
  we present results of a search for cold fronts in 
  high-resolution simulations of galaxy clusters in cold dark matter
  (CDM) models. We show that sharp gradients with properties similar
  to those of observed cold fronts naturally arise in cluster mergers
  when the shocks heat gas surrounding the merging sub-cluster, while
  its dense core remains relatively cold.  The compression induced by
  supersonic motions and shock heating during the merger enhance the
  amplitude of gas density and temperature gradients across the front.
  Our results indicate that cold fronts are non-equilibrium transient
  phenomena and can be observed for a period of less than a billion
  years.  We show that the velocity and density fields of gas
  surrounding the cold front can be very irregular which would
  complicate analyses aiming to put constraints on the physical
  conditions of the intracluster medium in the vicinity of the front.
\end{abstract}


]

\section{Introduction}
\label{sec:intro}

Recently, high-resolution {\sl Chandra} observations revealed the
existence of very sharp features in the X-ray surface brightness and
temperature maps of several clusters, including A2142
\citep{markevitch00} and A3667 \citep{vikhlinin01}. The sharp
gradients in the \xray surface brightness were first found in the {\sl
ROSAT} images of A2142 and A3667 along with signatures of
substructures (i.e., several small groups and merging components
embedded in a larger cluster).  Due to lack of spatially resolved
temperature maps, these features were initially interpreted as shock
fronts arising during cluster mergers \citep{markevitch99}.  This
interpretation turned out to be incorrect when detailed \xray surface
brightness and temperature maps were obtained with the {\sl Chandra}
satellite.  The maps showed that the increase in gas density and \xray
surface brightness by a factor of $\gtrsim 2$ across the sharp
features is accompanied by a {\em decrease} in temperature of gas of
a similar magnitude, the behavior opposite to that expected across
shock fronts.  These sharp features were therefore named {\em ``cold
fronts''}.  \citet{markevitch00} and \citet{vikhlinin01} interpreted
these fronts as a boundary of ``the dense subcluster core that has
survived a merger and ram pressure stripping by the surrounding
shock-heated gas.''

The existence of the sharp temperature gradients can be used to put
interesting constraints on the conditions in the intracluster medium
(ICM) in the vicinity of the front. In particular, the width of the
observed front ($\lesssim 5-10$~kpc) is several times smaller than the
Coulomb mean free path for electrons in the ICM.  This indicates that
the thermal conduction must be suppressed, at least across the
front. In the case of A2142, \citet{ettori00} find that the thermal
conductivity has to be reduced by several orders of magnitude from the
classical Spitzer value near the front.  The observed extent,
$\sim 500$~kpc, of the sharp boundary may also indicate that the
Kelvin-Helmholtz instabilities are partially suppressed, most likely
by magnetic fields parallel to the boundary.  The instabilities,
expected to arise on the boundary if the hot gas flows along the
front, would disrupt and widen the boundary.  Using a simple dynamical
model that approximates the gas cloud as a dense spherical cloud,
\citet{vikhlinin01} find that the magnetic field of 7-16$\mu$G
parallel to the front must be present at the boundary to maintain its
stability.  The detailed studies of cold fronts can, thus, provide new
detailed insights into the physics of the ICM, including the
efficiency of energy transport and the magnetic field strength in the 
ICM. The reliability of such models, however, depends on 
understanding the dynamics of gas motions during cluster mergers and,
especially, in the vicinity of the cold fronts. The constraints on the
transport mechanisms require also that we understand how long the cold
fronts can survive dynamically.

In this paper we present results of the search and analysis of
features similar to the observed cold fronts in very high-resolution
cosmological N-body$+$gasdynamics simulations of clusters forming in
Cold Dark Matter (CDM) models.  The main goals of this paper are 1) to
test the interpretation of the cold fronts as boundaries between hot
shock-heated gas and dense cold gas of the merging sub-clump; 2)
identify the situations and mechanisms that can produce cold fronts
and 3) analyze the dynamics and density fields of gas and dark matter
in the vicinity of the front.

Although numerous extensive theoretical analyses of cluster mergers
have been done in the past, the typical spatial resolution in studies
of cluster formation in the full cosmological context is too low to
match the superb spatial resolution of the {\sl Chandra} images.
Eulerian gasdynamics codes based on the modern high-resolution shock
capturing techniques are capable of resolving sharp density and
temperature gradients of arbitrary amplitude within $1-2$ grid cells
and are thus well suited for studies of shocks and cold fronts in
cluster mergers. However, high-dynamic range required in
self-consistent simulations of cluster formation limited use of the
Eulerian codes to controlled merger experiments
\citep[e.g.,][]{roettiger97a,quilis98,ricker01}.  The results of these
simulations were widely used for physical interpretation of the new
high-resolution \xray observations of merging clusters.  More
recently, the advent of cosmological codes using the adaptive mesh
refinement (AMR)
\citep[e.g.,][]{bryan97,kravtsov02,teyssier02}
allowed to achieve the dynamic range of $\gtrsim 10^4-10^5$ with
Eulerian gasdynamics algorithms, thus making possible 
high-resolution self-consistent cluster simulations in a realistic
cosmological setting.  As we will show below, these simulations can
resolve sharp gradients in density and temperature fields of the ICM
on scales of $\sim 10$~kpc, approaching the typical resolution of {\sl
Chandra} images.

The paper is organized as follows. In $\S$2, we discuss the cluster
simulations used in our analysis.  In $\S$3, we discuss and illustrate
the detailed structural evolution of the cluster gas and surrounding
filaments during cluster mergers and accretion of matter.  In \S~4 we
present cold fronts identified in numerical simulations and discuss
their origin, structure, dynamical properties, and detectability. We
discuss our results and summarize our conclusions in \S~5.

\begin{figure*}[t]
\centerline{ 
   \epsfysize=1.7truein  \epsffile{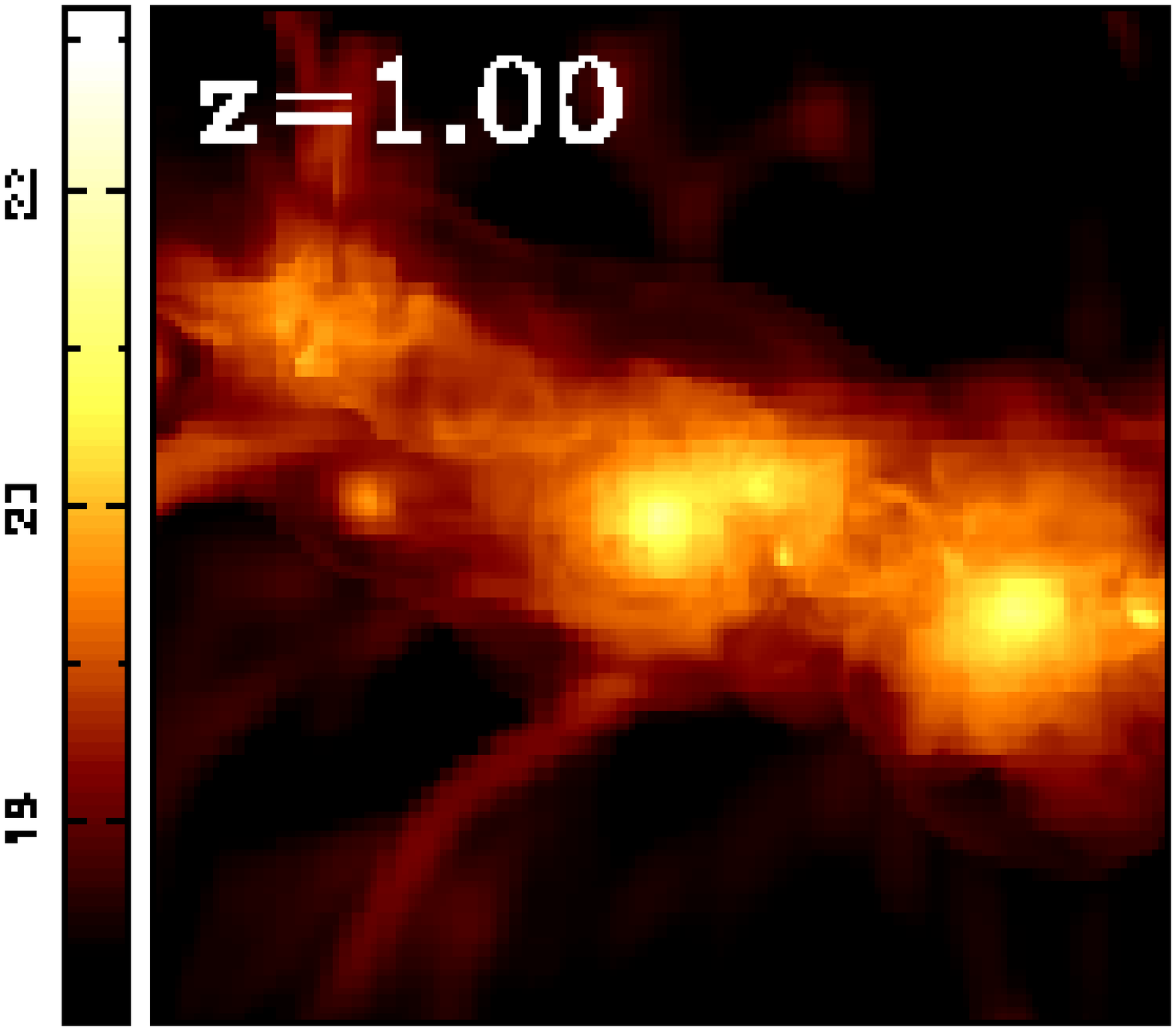}
   \epsfysize=1.7truein  \epsffile{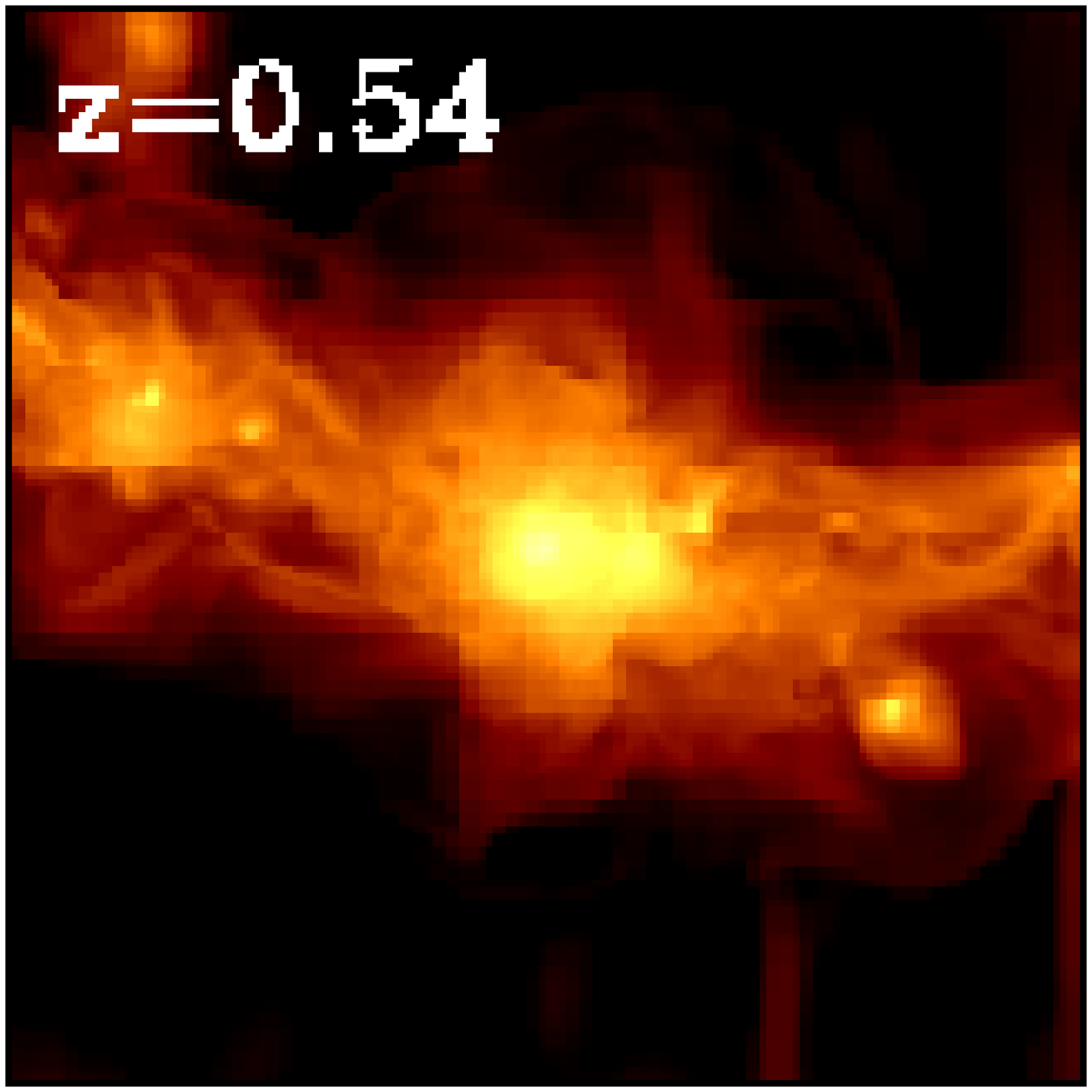}
   \epsfysize=1.7truein  \epsffile{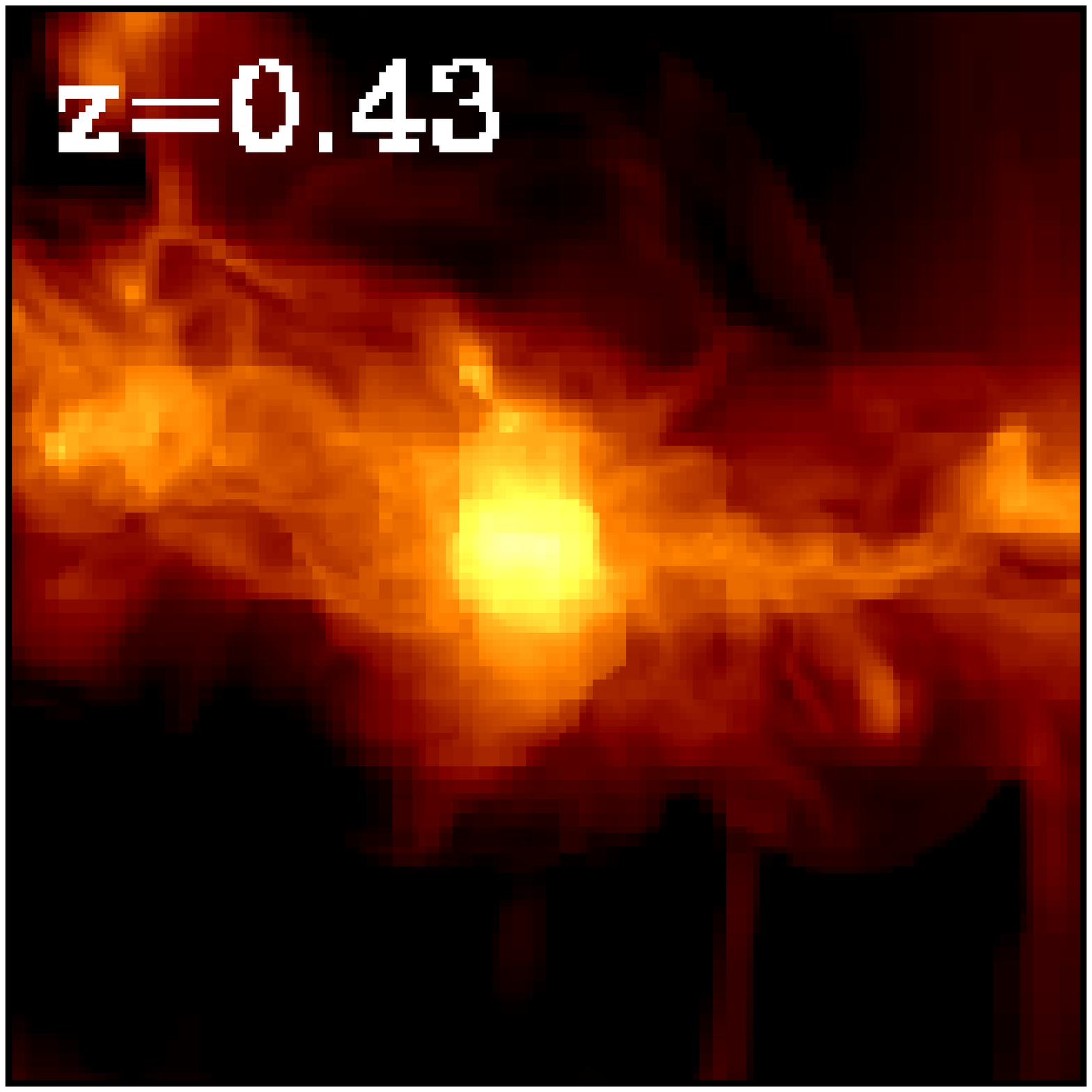}
   \epsfysize=1.7truein  \epsffile{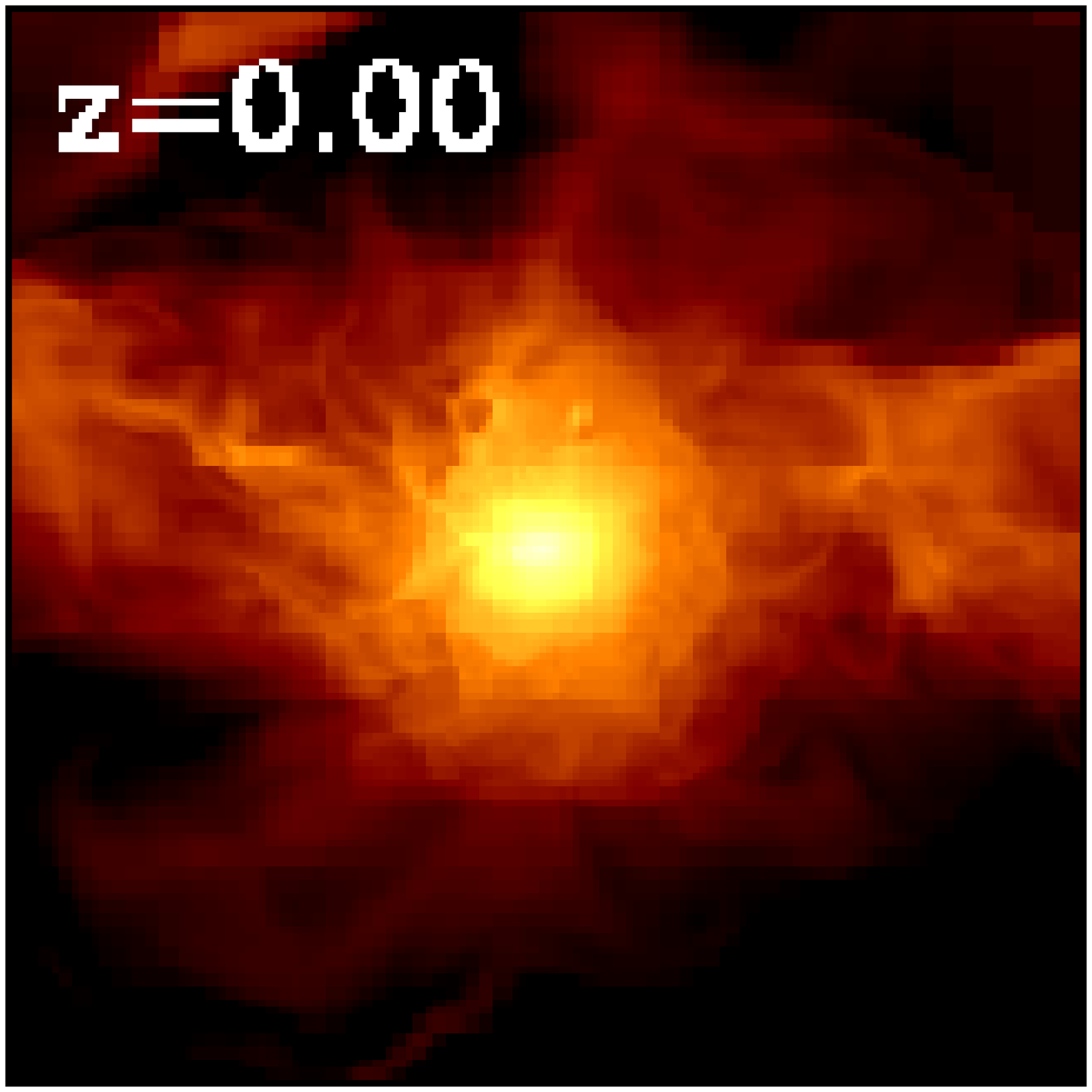}
}                                                  
                                                   
\centerline{ 
   \epsfysize=1.7truein  \epsffile{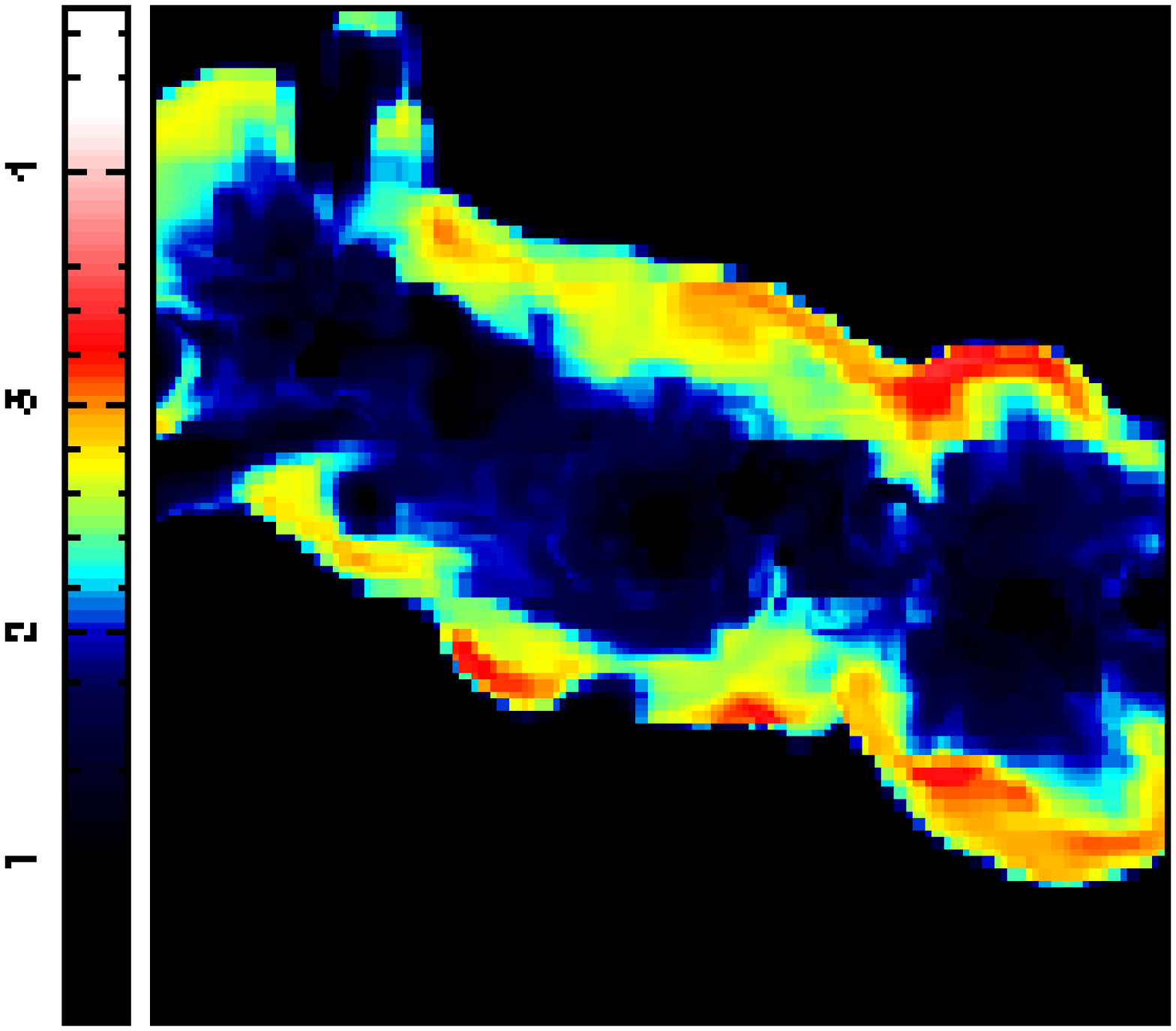}
   \epsfysize=1.7truein  \epsffile{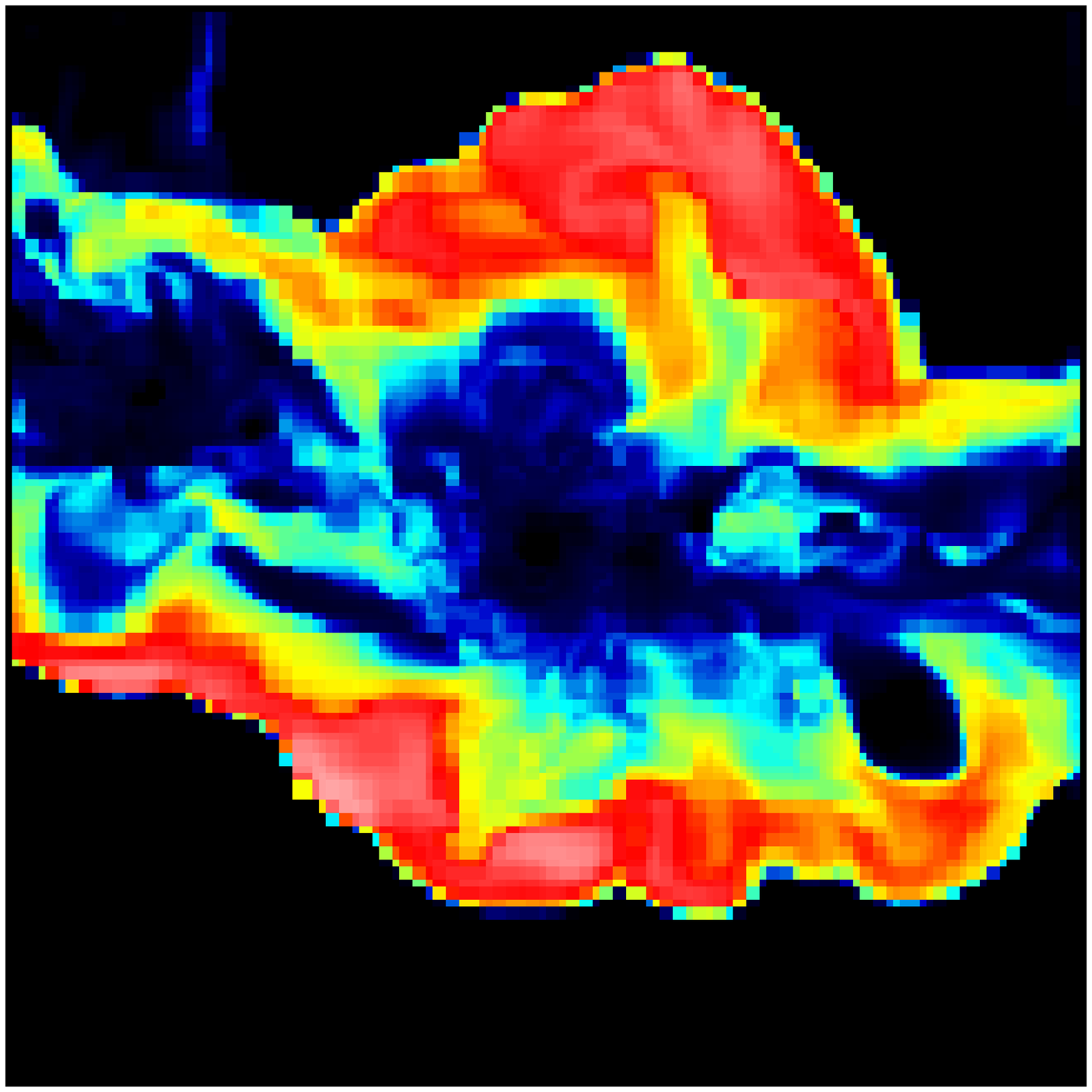}
   \epsfysize=1.7truein  \epsffile{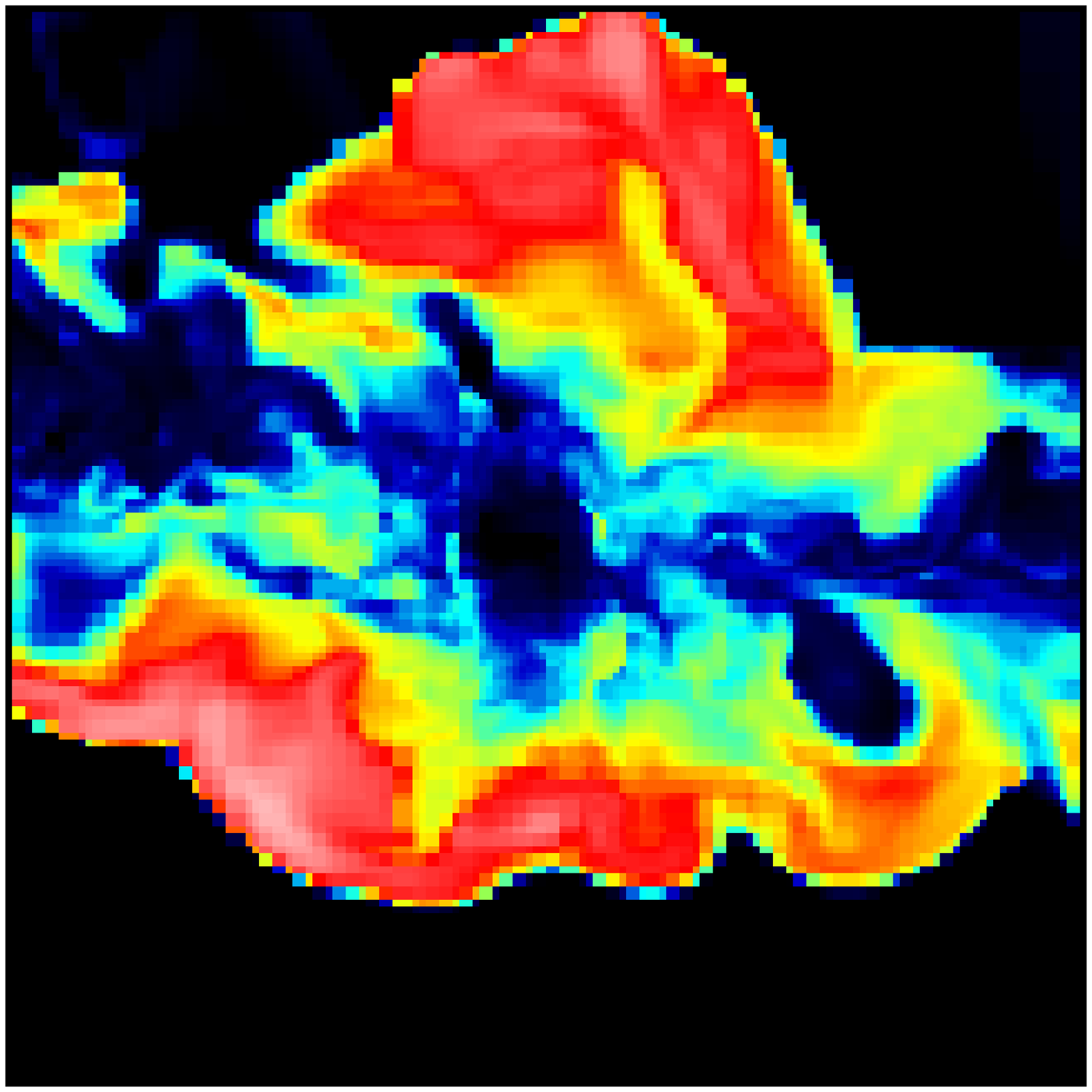}
   \epsfysize=1.7truein  \epsffile{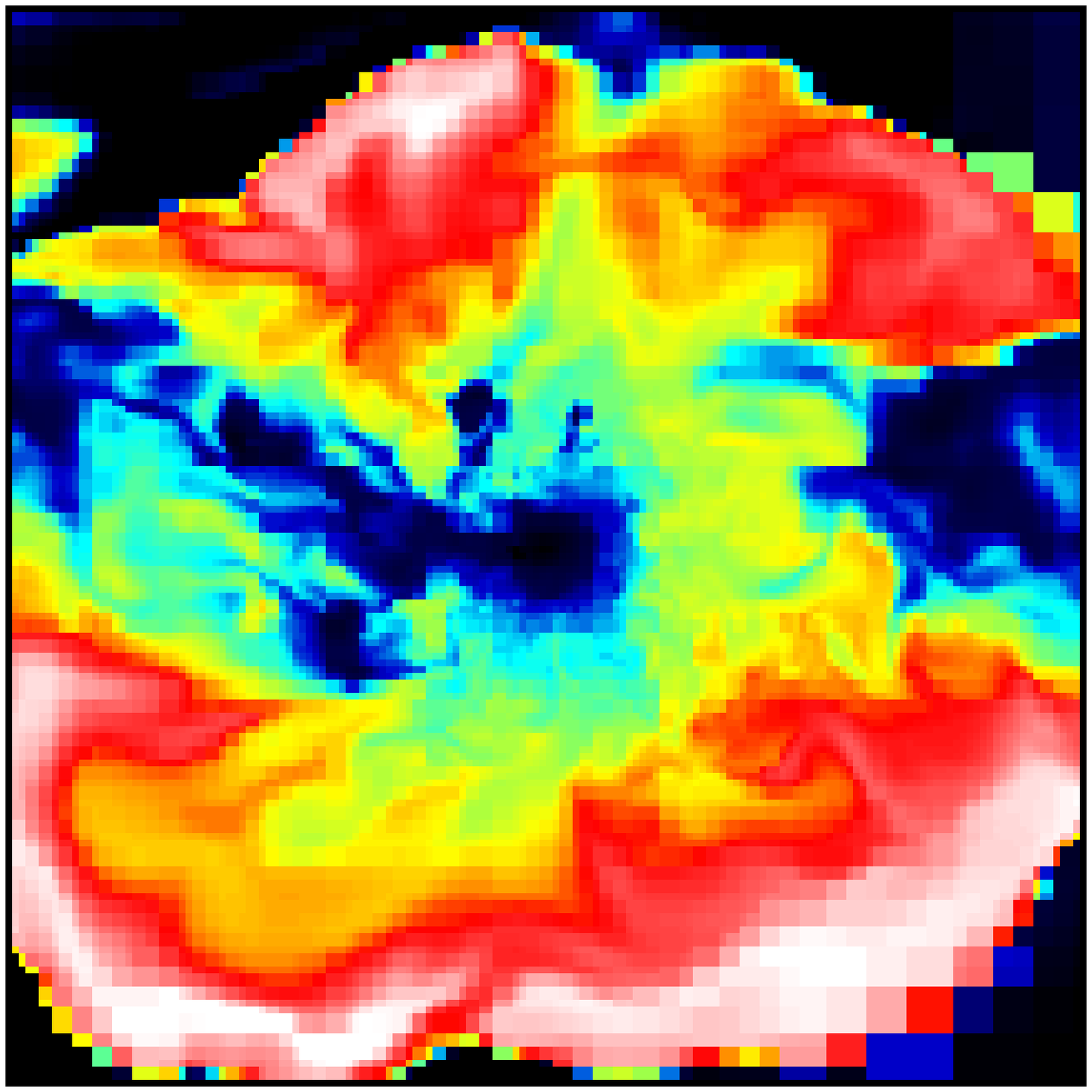}
}
\caption{ \footnotesize
The maps of projected density (top) and entropy (defined as
$Tn^{-2/3}$, bottom) of a simulated $\Lambda$CDM cluster at four
different redshifts in a $60h^{-1}$~kpc slice centered on the central
density peak. The maps are color-coded on a $\log_{10}$ scale in units
of cm$^2$ (density) and keV cm$^2$ (entropy). The size of the 
region shown is 8$h^{-1}$Mpc (10\% of the entire simulation volume). The
entropy maps reveal a very complex entropy distribution of the
gas. Both the filaments and the forming cluster are surrounded by
strong accretion shocks.  Note, however, that the accretion shock
around the cluster is very aspherical and does not penetrate into the
filament; relatively low-entropy gas accreting onto cluster along the
filament does not pass through the strong virial shocks and can be
traced all the way to the central $0.5\hMpc$ of the cluster. }
\label{clevol}
\end{figure*}

\section{Numerical simulations}
\label{sec:numdetail}

In this study, we analyze the high-resolution cluster simulations
performed using the Adaptive Refinement Tree (ART)
$N$-body$+$gasdynamics code \citep{kravtsov99, kravtsov02}.  ART
is an Eulerian code designed to achieve high spatial resolution by
adaptively refining regions of interest, such as high-density regions
or regions of steep gradients in gas properties, and has good
shock-capturing characteristics.  The code can capture discontinuities
in gas properties within $\sim 1-2$ grid cells, which makes it
well-suited for studying sharp features such as shock and cold fronts.
At the same time, the adaptive refinement in space and (non-adaptive)
refinement in mass \citep{klypin01} allows us to reach the high
dynamic range required for high-resolution self-consistent simulations
of cluster evolution in a cosmological setting.

We note that the effects of magnetic fields, gas cooling, stellar
feedback, and thermal conduction are not included in these
simulations.  This is the first theoretical study of cold fronts and
it is therefore interesting to see whether these features can be
produced in the case of purely adiabatic gasdynamics. As we show
below, cold fronts arise at the interfaces between relative low
density hot medium and dense cold gas of merging clumps.  These
regions should not be significantly affected by gas cooling, which is
expected to have the largest effect in cluster cores. The absence of
thermal conduction is acceptable because our primary goal is to study
the dynamical origin of cold fronts, rather than their precise
properties. The very existence of cold fronts indicates that thermal
conduction is suppressed in their vicinity.  Other theoretical
arguments \citep[e.g.,][]{loeb02} constrain the thermal conduction
coefficient to be significantly smaller than the classical Spitzer
value over most of the ICM, except perhaps in the cluster cores
\citep{voigt02}.  Our simulations, in fact, implicitly assume that the
Coulomb mean free path is smaller than the size of computational cells
(this is one of the main assumption of gasdynamics), and hence there
is no thermal conduction.

We analyze simulations of two clusters of intermediate mass in the
$\Lambda$CDM and standard CDM models.  The first ($\Lambda$CDM)
cluster was simulated in the ``concordance'' flat {$\Lambda$}CDM
model: $\Omega_m=1-\Omega_{\Lambda}=0.3$, $\Omega_b=0.043$, $h=0.7$
and $\sigma_8=0.9$, where the Hubble constant is defined as $100h{\
\rm km\ s^{-1}\ Mpc^{-1}}$, and $\sigma_8$ is the power spectrum
normalization on $8h^{-1}$~Mpc scale. The simulation used 128$^3$
uniform grid and 7 levels of mesh refinement in the computational box
of $80h^{-1}$~Mpc, which corresponds to the dynamic range of
$128\times 2^7=16,384$ and peak resolution of $80/16,384\approx
5\hkpc$. Only the region of $\sim 10$\hmpc~ around the cluster was
adaptively refined, the rest of the volume was followed on the uniform
$128^3$ grid. Similarly, the simulation used initial conditions
sampled with different particle masses \citep[see
][]{klypin01}. The mass resolution (dark matter particle mass) in
the region around the cluster was $2.7\times 10^{8}h^{-1}{\rm\
M_{\odot}}$, while other regions were simulated with lower mass
resolution (the total of three particle species, different in mass by
a factor of eight, were used). The region of the cluster was selected
for re-simulation at higher resolution from a lower resolution
$N$-body run.

The second (SCDM) cluster is simulated from the initial conditions
used in the ``Santa Barbara cluster comparison project''
\citep{frenk99}.  The cosmological parameters assumed were those of
the standard CDM cosmology: $\Omega_m=1.0$, $\Omega_b=0.1$, $h=0.5$
and $\sigma_8=0.9$.  The simulation used 128$^3$
uniform grid and 6 levels of refinement, which corresponds to the peak
resolution of $\approx 4h^{-1}$~kpc in the $32h^{-1}$~Mpc
computational box. Here, again we used a resampling technique to set up
the initial conditions for multiple particle species.  The lowest particle
mass is $4.9\times 10^{8}h^{-1}{\rm\ M_{\odot}}$ and corresponds to
$256^3$ effective number of particles. The Lagrangian region
corresponding to the spherical volume of the radius $R=5R_{\rm vir}$,
where $R_{\rm vir}$ is the virial radius of the cluster at $z=0$, was
simulated with the highest mass resolution and was adaptively refined.
The SCDM cluster simulation was previously used to compare performance
of the ART code to other existing cosmological gasdynamics codes in a
realistic cosmological setting \citep{kravtsov02}, where good
agreement was found in both cluster properties and its radial
profiles.

The SCDM and $\Lambda$CDM clusters had virial masses of
$M_{200}=6.0\times 10^{14}\hmsol$ and $M_{340}=2.4\times
10^{14}\hmsol$ and virial radii of $R_{200}=1.35$\hmpc~ and
$R_{340}=1.26$\hmpc, respectively.  Here subscripts indicate the
cosmology-dependent virial overdensity with respect to the mean
density of the universe. Both clusters were therefore simulated with
about one million dark matter particles within the virial radius. In
addition to $128^3$ cells of the uniform grid, approximately $10^7$
mesh cells were generated in the process of adaptive refinement in the
regions surrounding clusters.  The simulations were thus designed to
concentrate the bulk of the computational effort on the individual
clusters, which allowed the simulations to reach the high dynamic range
critical for this study. The spatial resolution of these simulations
allows us to resolve sharp gradients over scales $\lesssim 20\hkpc$
within the virial radius of clusters. This is comparable to, albeit
still somewhat worse than, the superb angular resolution of the {\sl
Chandra} satellite.

\section{Cluster Evolution}
\label{sec:evolution}

Let us start the discussion of cold fronts with a general description of
cluster evolution.  Figure~\ref{clevol} shows maps of projected gas
density (top) and entropy (defines as $T n^{-2/3}$, bottom) of the
simulated $\Lambda$CDM cluster at four different redshifts in a
$60h^{-1}$~kpc slice centered on the central gas density peak. The
maps are color-coded on a $\log_{10}$ scale in units of cm$^2$
(density) and keV cm$^2$ (entropy). The size of the region shown is
8$h^{-1}$Mpc (10\% of the entire simulation volume).  The figure
illustrates complex dynamical processes accompanying cluster
formation: accretion of clumps and diffuse gas along filaments, strong
accretion and virial shocks both around cluster and surrounding
filaments, weaker merger shocks within the virial shock of the
cluster, and the complicated flow pattern of the ICM gas revealed by
the entropy map.

The evolution of intergalactic gas is driven by accretion and mergers
between clumps.  Between $z=1$ and $z=0.5$, the main cluster undergoes
a nearly equal mass merger.  Such mergers are spectacular events
involving the kinetic energy as large as $\sim 10^{64}$ ergs, the most
energetic events since the Big Bang.  More common are smaller mergers
and accretion of groups along large-scale filaments. At $z\sim 1-3$,
highly supersonic accretion (Mach numbers of $\gtrsim 100$) of
pristine gas from low-density regions leads to formation of strong
shocks located roughly at the virial radii of the two merging
proto-clusters.  Note that shocks of similar strength also surround
the filament along which these clusters move. Note also that virial
shocks do not penetrate into the filament itself and it is in fact
difficult to separate the virial shocks of clusters from the accretion
shocks around the filament.

During the evolution from $z=1$ to $z=0$, as the mass of the main
cluster grows, the radius and entropy gradient of the virial shocks
steadily increase. Similar increases can be observed for the accretion
shocks around the filament.  However, even at the present epoch the
gas flowing along the filament reaches the central regions of cluster
without passing through a strong virial shock. This is clearly
illustrated in the entropy maps in which relatively low-entropy gas
flowing along the filament can be traced all the way to the central
$0.5\hMpc$ of the cluster.  The virial shocks instead propagate along
directions of the steepest pressure gradient into the low-density
voids.  The flow of filamentary gas, with entropy already increased by
the accretion shocks (temperature of several million degrees K), has
considerably smaller Mach numbers (typically $\sim 1-10$).  Its motion
is therefore slightly supersonic at best, compared to the highly
supersonic motion of the yet unshocked gas flowing from voids. When
this gas reaches the cluster core it accelerates and generates random,
weak shocks in which it dissipates its kinetic energy. The flows from
different filaments and directions result in a complicated flow
patterns. We will discuss these motions in relation to the cold fronts
below (see also Figure~\ref{dm_temp}).

\begin{figure*}[t]
\centerline{ 
   \epsfysize=1.7truein  \epsffile{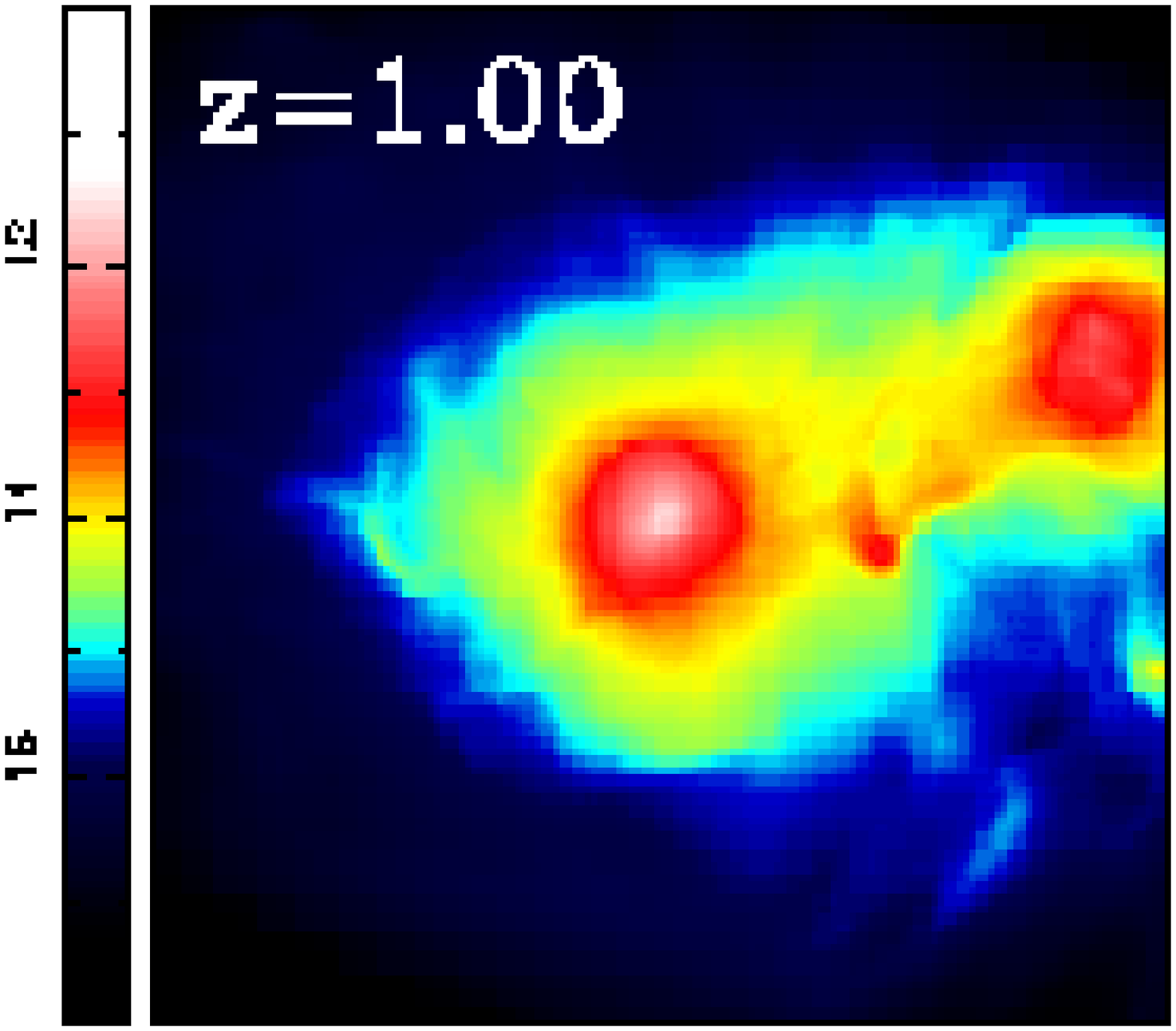}
   \epsfysize=1.7truein  \epsffile{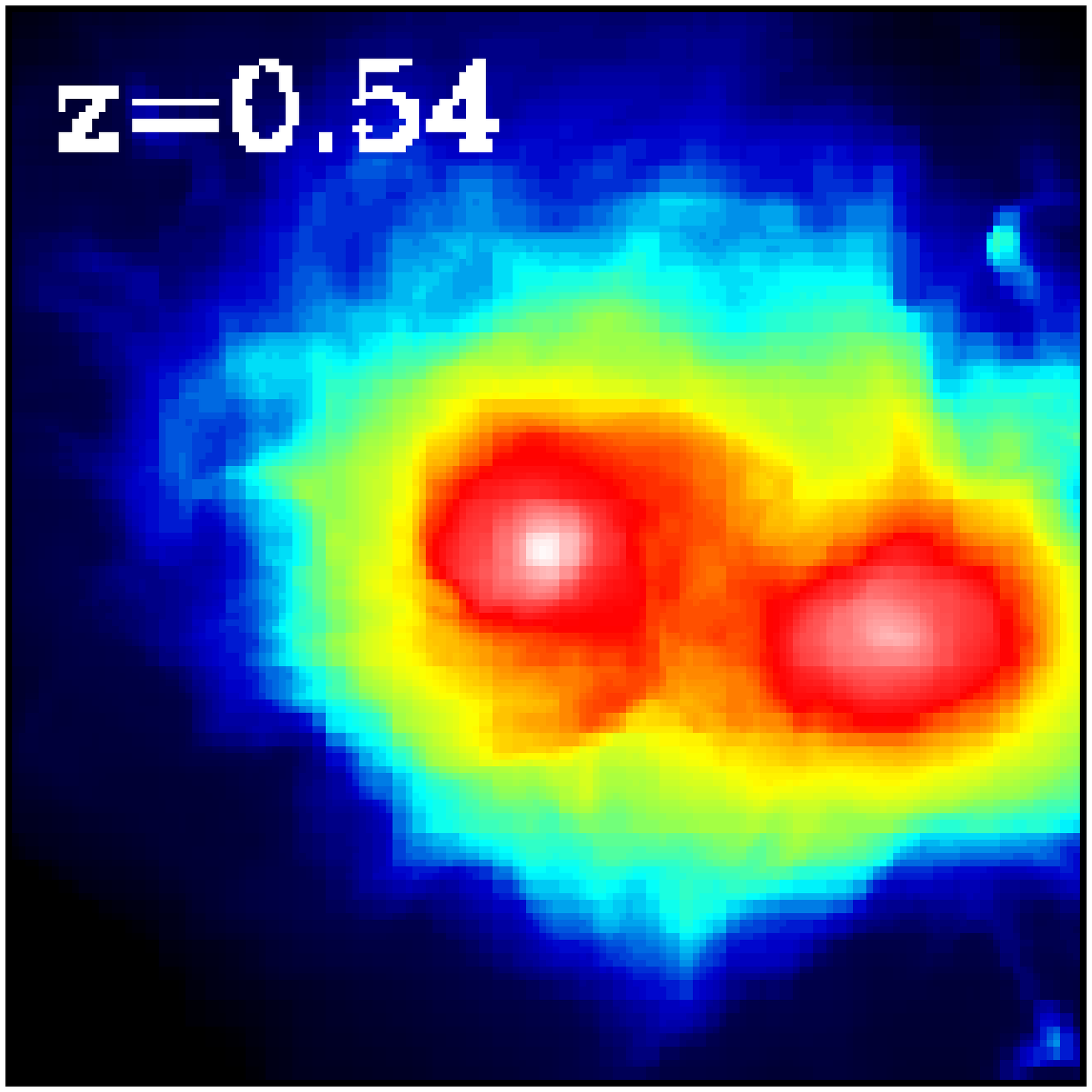}
   \epsfysize=1.7truein  \epsffile{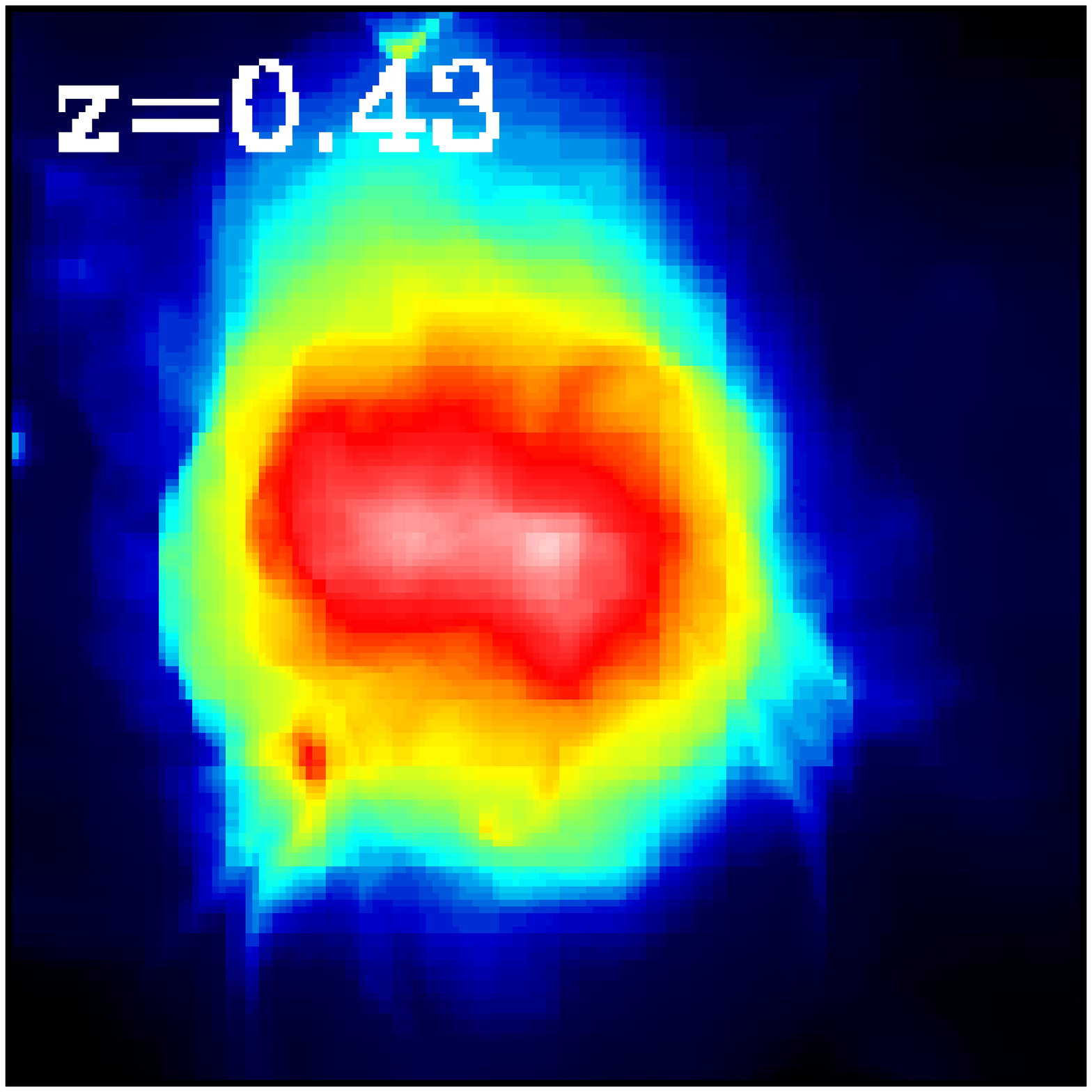}
   \epsfysize=1.7truein  \epsffile{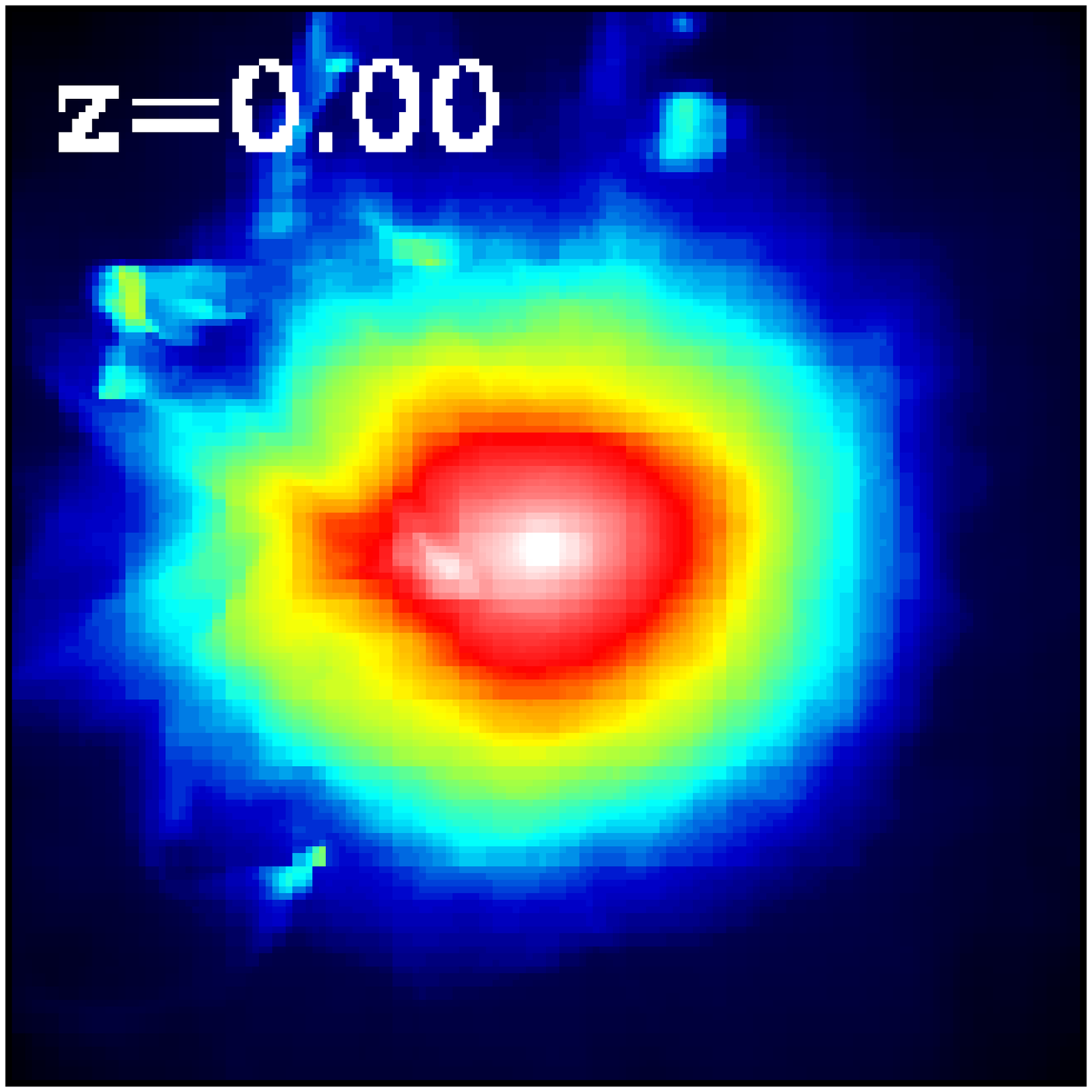}
}                                                  
                                                   
\centerline{ 
   \epsfysize=1.7truein  \epsffile{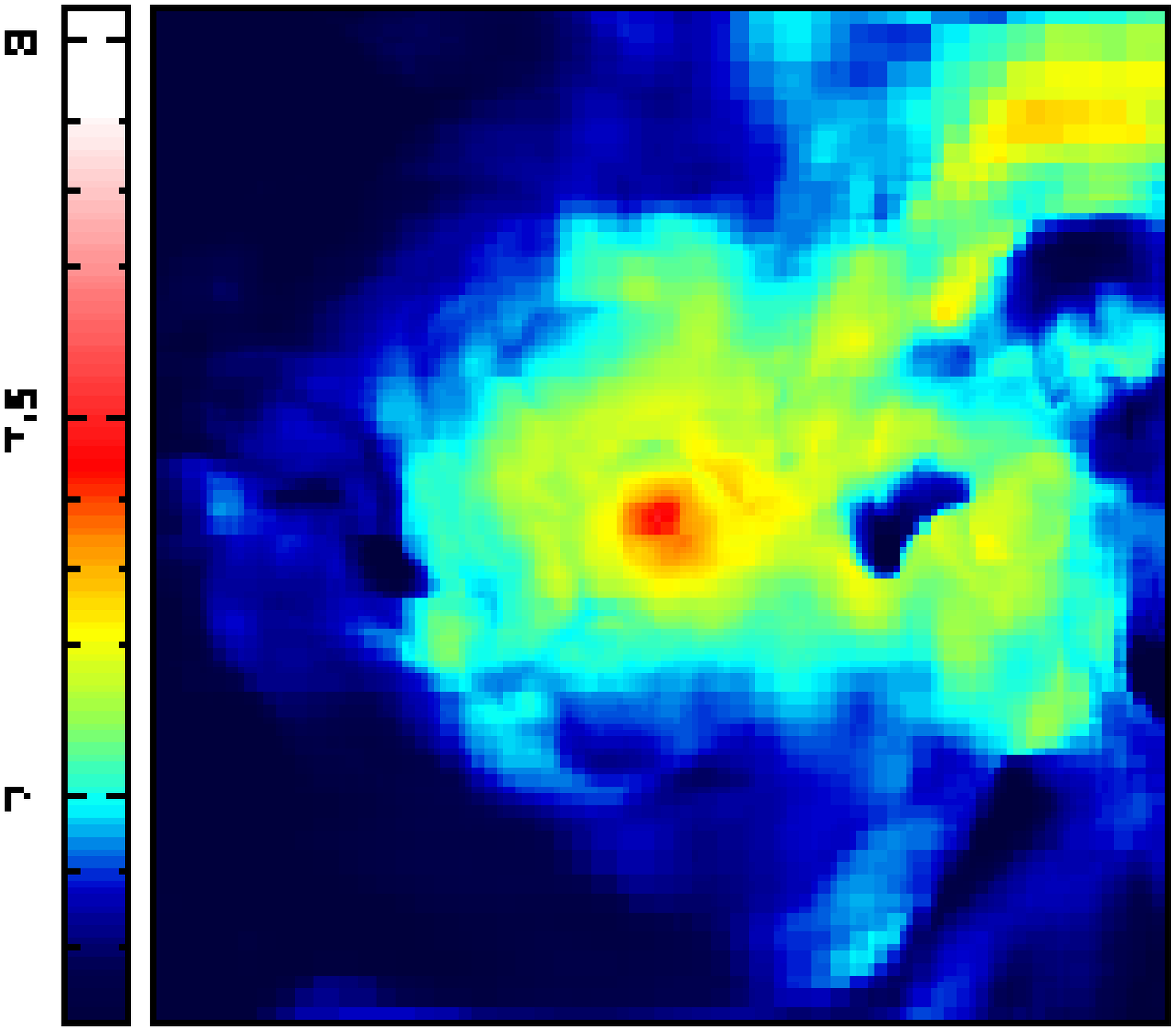}
   \epsfysize=1.7truein  \epsffile{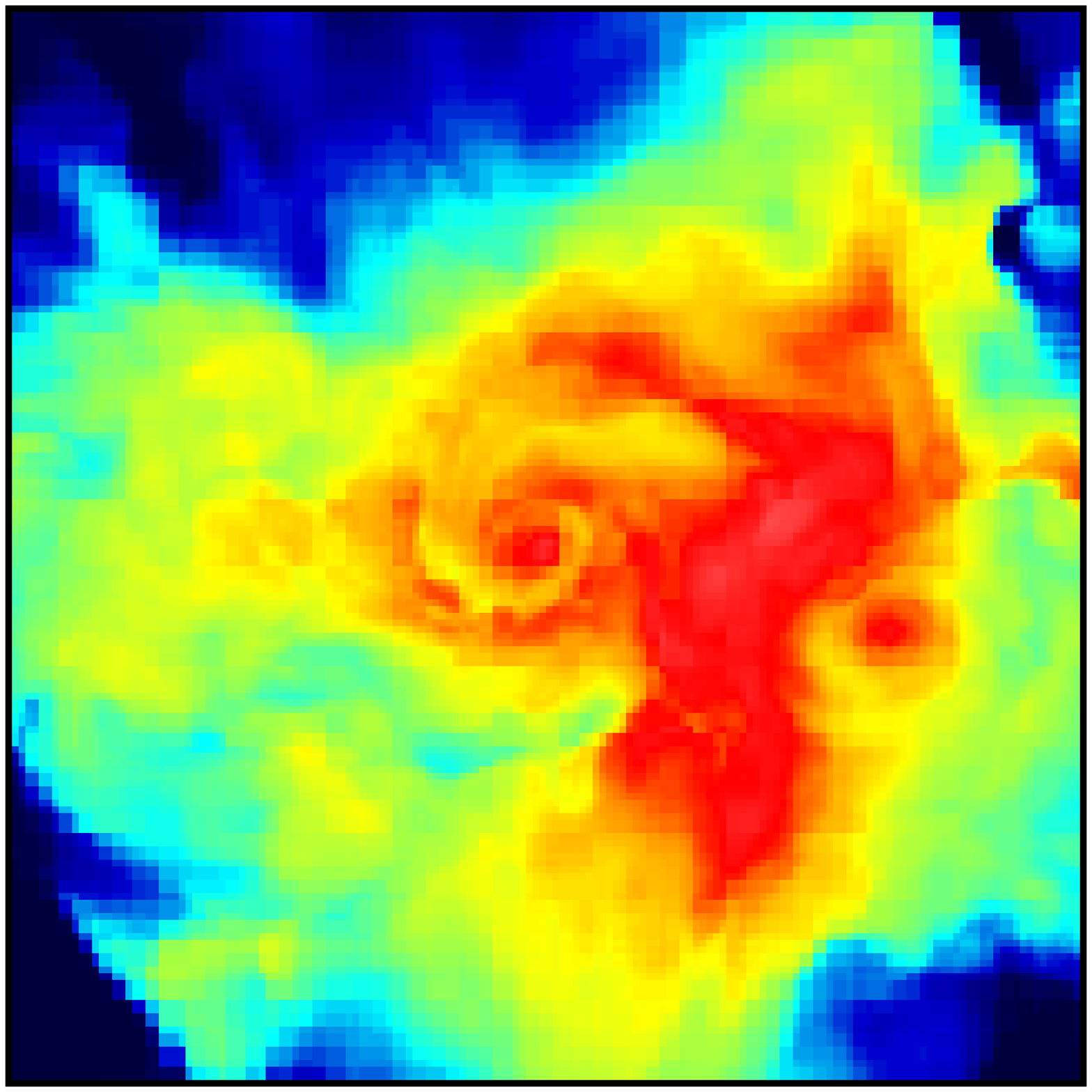}
   \epsfysize=1.7truein  \epsffile{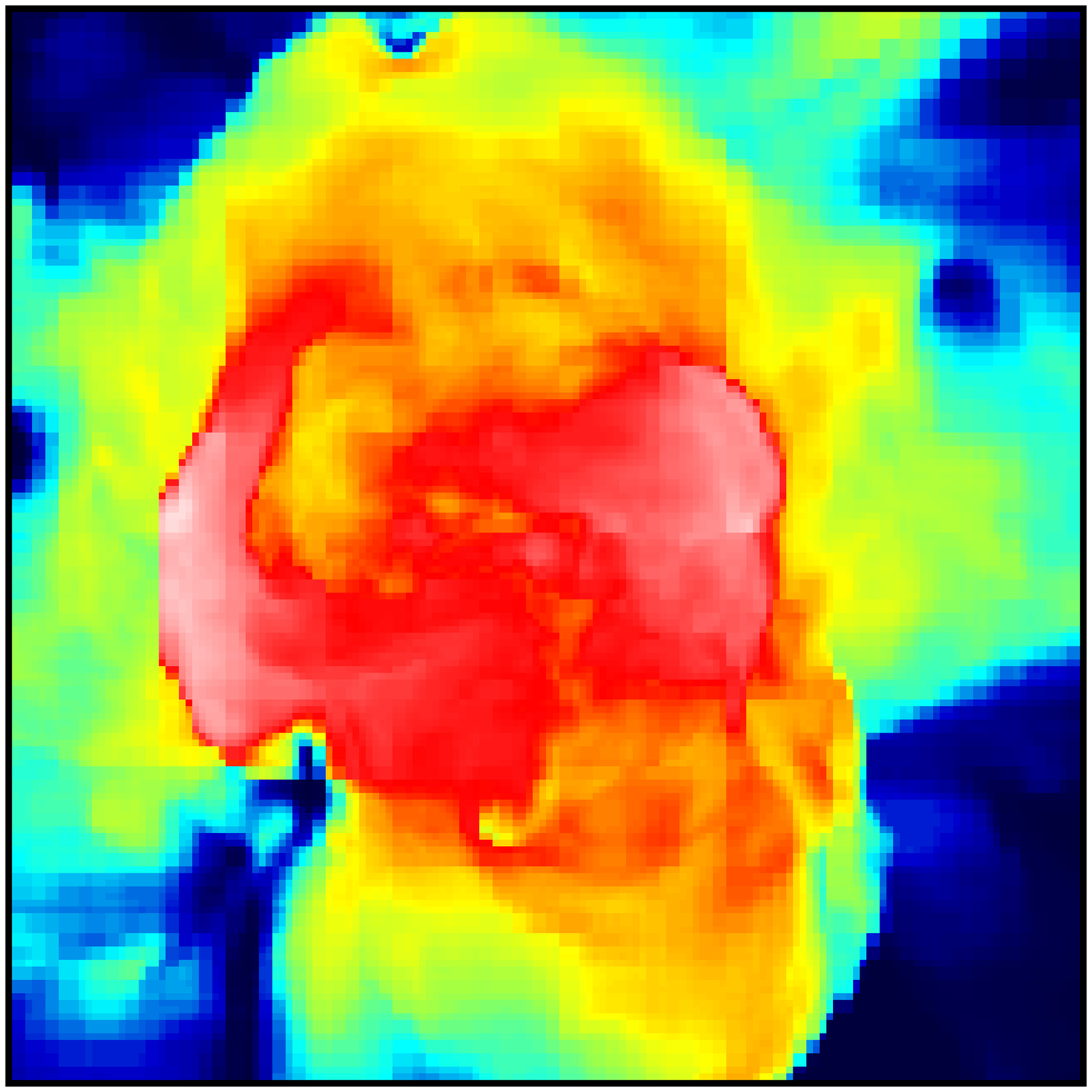}
   \epsfysize=1.7truein  \epsffile{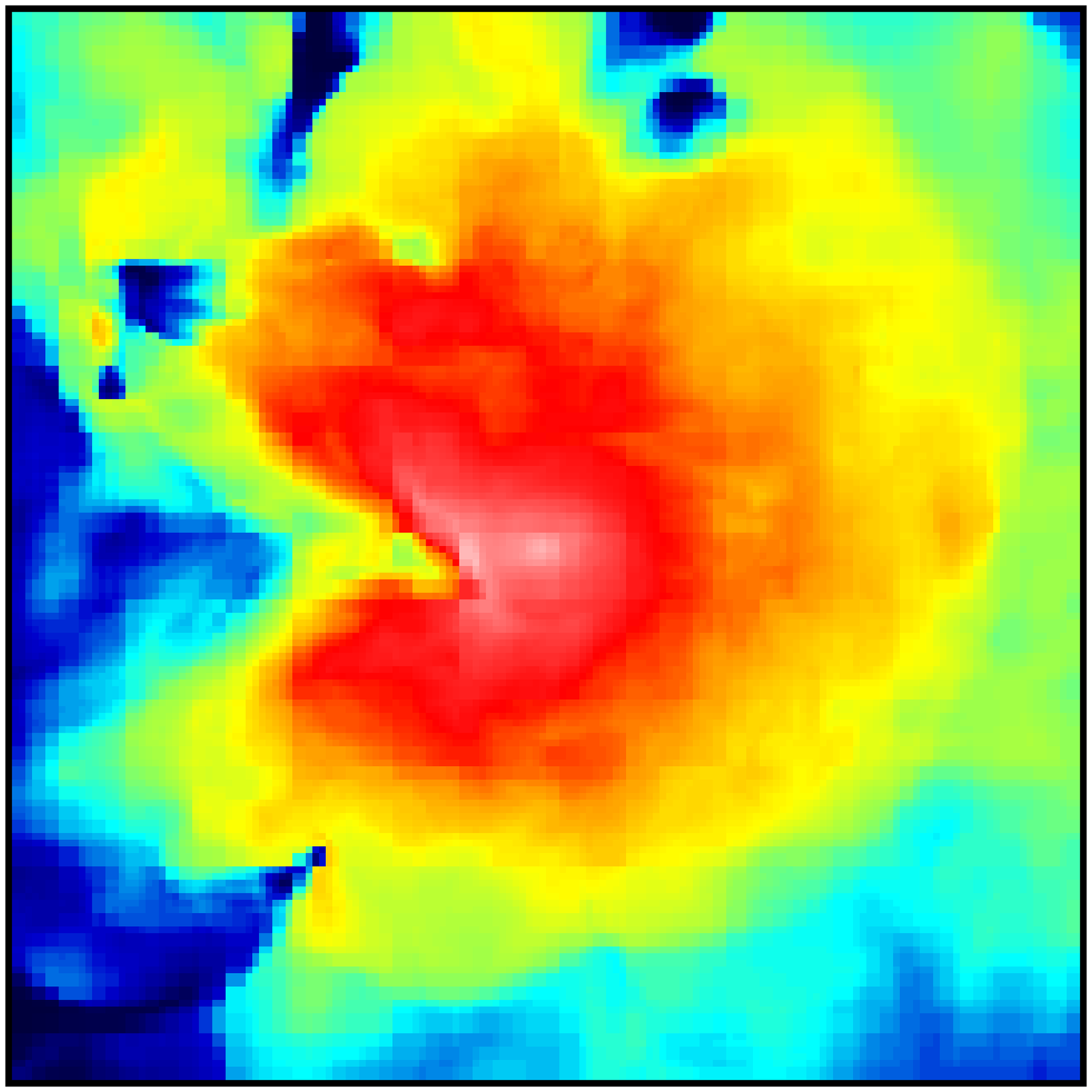}
}
\caption{ \footnotesize
The X-ray surface brightness (top) and emission-weighted temperature
(bottom) of the simulated $\Lambda$CDM cluster at four different
redshifts.  The maps are color-coded on a $\log_{10}$ scale in units of
erg~s$^{-1}$~cm$^{-2}$~arcmin$^{-2}$ (surface brightness) and keV
(temperature).  Both surface brightness and temperature are calculated
in the 0.5-2~keV band.  The size of the volume shown is 2$h^{-1}$
Mpc. Note the cold front associated with the merging sub-cluster
behind the merger shock front to the left of two sub-clusters in the
$z=0.43$ panel.  Note also the cold front associated with the merging
subclump at $z=0$ to the left of the cluster center and the adjacent
compressed region of enhanced temperature. The subclump is trailed by
a relatively cold ($\sim$1-2~keV) intergalactic gas accreted along a
filament. }
\label{xsbt_map}
\end{figure*}

\section{Cold fronts}
\label{sec:coldfront}

The main question we want to answer in this study is whether
cosmological simulations of galaxy clusters forming in hierarchical
models can produce features similar to the ``cold fronts'' observed by
the {\sl Chandra} satellite. If so, simulations can then be used to
shed light on the origin and properties of these features.  In this
section, we analyze the high-resolution cluster simulations described
above to search for sharp gradients in temperature and density
structure of clusters.  

\placefigure{xsbt_map}

\subsection{X-ray surface brightness and temperature maps}
\label{sec:xray}

To perform the search and analysis of cold fronts, we generate the
X-ray surface brightness and emission-weighted temperature maps of the
simulated clusters. The maps were generated from uniform density and
temperature grids centered on the minimum of cluster potential.  The
density and temperature grids, in turn, were constructed using
interpolation from the original mesh refinement structure. The X-ray
emissivity of gas cells was calculated in the 0.5-2~keV energy band
using the
\citet{raymond77} plasma emission model and assuming a uniform
metallicity of 0.3 solar.  Figure~\ref{xsbt_map} shows the X-ray
surface brightness (top) and emission-weighted temperature (bottom)
maps of the simulated $\Lambda$CDM cluster at four different
redshifts.  The maps are color-coded on a $\log_{10}$ scale in units
of erg~s$^{-1}$~cm$^{-2}$~arcmin$^{-2}$ (surface brightness) and keV
(temperature). The region shown in the figure is 2$h^{-1}$ Mpc; the
pixel size is $2\hMpc/256\approx 8\hkpc$. The size of the pixels were
chosen to be similar to the effective resolution of the simulation
within the cluster.  

These maps show that complex density and temperature structure of the
intracluster gas, illustrated by the thin density and entropy slices
in Figure~\ref{clevol}, is present in the observables averaged over
the line-of-sight. At $z=1$, two clusters are approaching each other
along a massive filament.  These clusters undergo a nearly equal-mass,
slightly off-axis merger at $z=0.54$.  Compression of gas between two
merging clumps significantly enhances the temperature in the large
 band visible in $z=0.54$ panel. By $z=0.43$, the clusters have passed
each other once and their supersonic motion results in two large-scale
bow shocks propagating in opposite directions along the merger axis.
These rather strong shocks are clearly seen in the temperature map of
$z=0.43$ epoch as regions of enhanced temperature with very sharp
boundaries.  At $z=0$, the cluster formed by the $z\sim 0.5$ merger is
relatively relaxed but continues to accrete matter and undergo minor
mergers (note a small merging sub-clump to the left of the cluster
center).

\subsection{Origin, properties and detectability}
\label{sec:properties}

The maps in Figure~\ref{xsbt_map} show that a variety of very sharp
features in temperature and surface brightness arises during minor
and, especially, major mergers.  Some of these features, indeed,
exhibit properties qualitatively similar to the observed cold fronts.
Below, we present two cases of cold fronts found in the CDM
simulations.

\subsubsection{Large-scale cold front during a major merger}
\label{sec:large}

\begin{figure}[t]
\centerline{ 
   \epsfysize=4.0truein  \epsffile{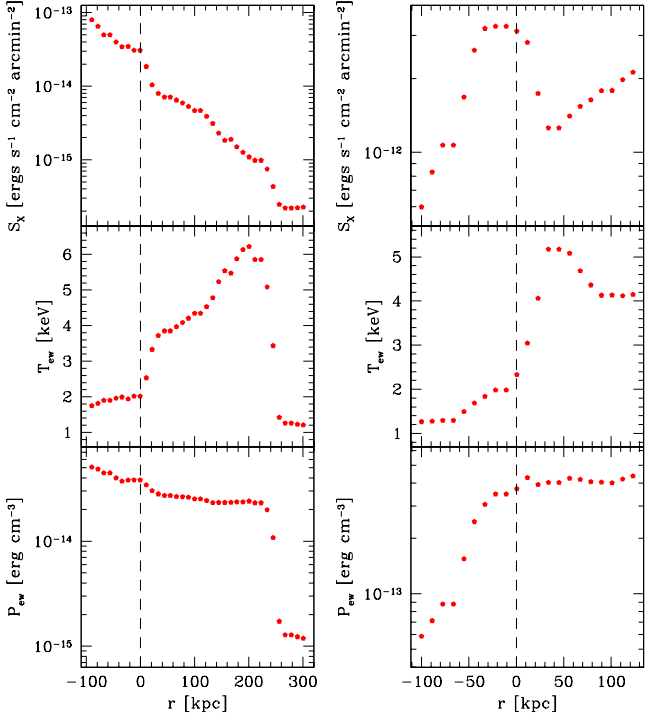}
}                                                  
\caption{ \footnotesize
  The X-ray surface brightness (top), emission-weighted temperature
  (middle) and emission-weighted pressure (bottom) profiles across
  cold fronts identified in z=0.43 (left) and z=0 (right) panels in
  the Figure 2.  The distance in kpc is measured relative to the cold
  fronts shown by dashed lines.  In both cases, a jump in surface
  brightness is accompanied by a drop in temperature over the scale of
  $\sim 30$~kpc, while pressure changes relatively smoothly across
  these features.  The simulated profiles and images reproduce all the
  main features of the observed cold fronts. The profiles in the left
  panels also show the presence of a bow shock $\approx 200$~kpc from
  the cold front. }
\label{xsbt_pro}
\end{figure}

\begin{figure}[t]
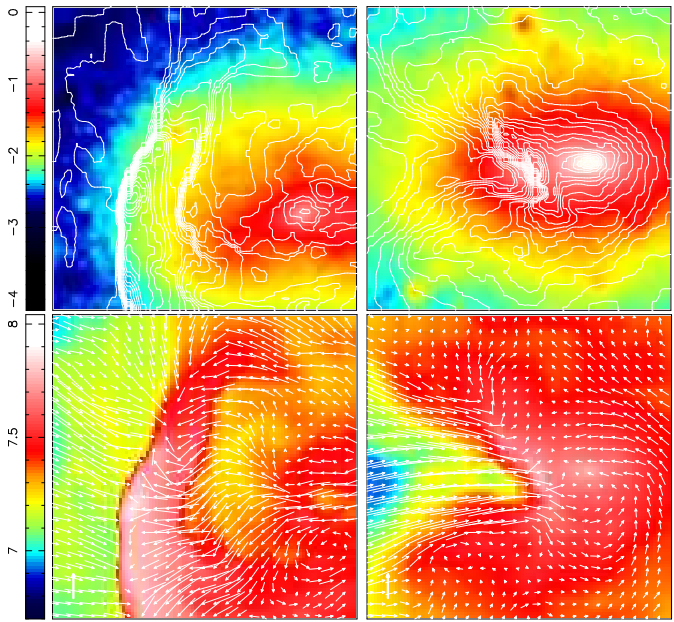

\centerline{
  \epsfysize=1.6truein  \epsffile{Plot_LowerRes/fig5a.ps}
  \epsfysize=1.6truein  \epsffile{Plot_LowerRes/fig5b.ps}
}                    
\centerline{    
  \epsfysize=1.6truein  \epsffile{Plot_LowerRes/fig5c.ps}
  \epsfysize=1.6truein  \epsffile{Plot_LowerRes/fig5d.ps}
}  
\caption{ \footnotesize
  {\it Top panels:\/} the map of projected dark matter density with
  overlaid contours of the emission-weighted gas temperature near the
  cold fronts identified in the $z=0.43$ (left) and $z=0$ (right)
  panels in Figure~\ref{xsbt_map}.  The dark matter density maps are
  color-coded on a $\log_{10}$ scale in units of g~cm$^{-2}$; the
  temperature contours are spaced on linear scale.  {\em Bottom
  panels:} the velocity maps of gas are overlayed on the
  emission-weighted temperature map of the regions shown in the top
  panels.  The length of the thick vertical vector in the bottom-left
  corner corresponds to 1000 \kms.  The size of the shown region is
  0.78$h^{-1}$Mpc.  The velocity maps show that the cold fronts are
  non-equilibrium phenomena arising during cluster mergers. }
\label{dm_temp}
\end{figure}

A spectacular cold front arises in a major merger of the $\Lambda$CDM
cluster at $z\sim 0.4-0.5$.  This cold front is a low temperature
region embedded in the high temperature gas just behind the left bow
shock in the $z=0.43$ panel of Fig.~\ref{xsbt_map}. The colder gas
belongs to one of the merging clusters. The dark matter halo of this
cluster is in the process of disruption and the cluster gas appears to
be ``sloshed out'' of its potential well (see Fig.~\ref{dm_temp}).
Rapid motion of the gas inside the freshly shocked gas then results in
a sharp boundary between cold unshocked gas of the cluster and the hot
shocked surrounding gas.  The left panels of Figure~\ref{xsbt_pro}
show the X-ray surface brightness (top), emission-weighted temperature
(middle) and emission-weighted pressure (bottom) profiles across this
temperature gradient in the $z=0.43$ maps.  The profiles were
constructed by drawing a straight line across the front and selecting
values of map pixels nearest to the line in equally spaced intervals.
The distance in kpc is measured relative to the sharp temperature
gradient indicated by the dashed line.  The profiles show that a drop
of temperature over the scale of $20-30\hkpc$ is accompanied by a jump
in surface brightness by a factor of $\gtrsim 2$, the behavior is very
similar to the profiles of observed cold fronts.  Despite jumps in gas
density and temperature, pressure changes relatively smoothly across
this boundary. The smooth change of pressure was also deduced for
observed cold fronts. Note that the profiles show the presence of a
bow shock $\approx 200$~kpc from the cold front.  A similar bow shock
was also observed in A3667 ahead of the cold front, although its
detection was marginal.

\begin{figure}[t]
\centerline{
  \epsfysize=3.0truein  \epsffile{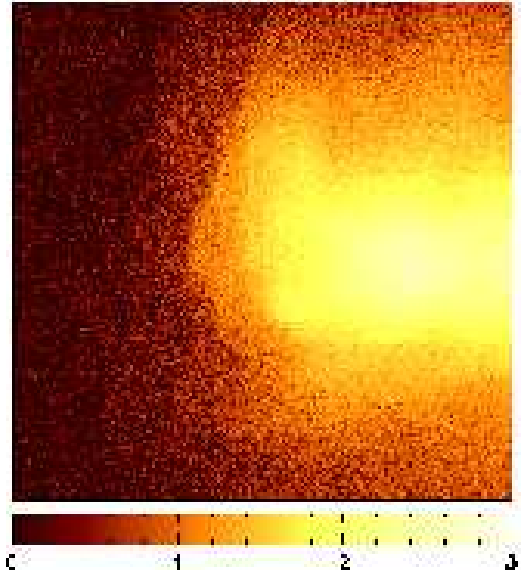}
}
\caption{ \footnotesize
  The mock {\sl Chandra} photon count image of the cold front visible
  in the $z=0.43$ panel of Figure~\ref{xsbt_map}. The image was
  constructed assuming 50~ksec exposure in the $0.5-4$ keV band,
  cluster redshift of $z=0.05$, and a flat background level of
  $4.3\times10^{-6}$~cnts~s$^{-1}$~pixel$^{-1}$.  The figure shows the
  full $16^\prime\times 16^\prime$ ACIS-I field of view binned to
  $2^{\prime\prime}$ pixels. The color represents photon counts per
  pixel in log$_{10}$ scale. } 
\label{chandra_map}
\end{figure}

The overall structure of the front also exhibits qualitatively similar
behavior to those observed in A3667 \citep{vikhlinin01}.  To
illustrate its appearance in a realistic observations, we constructed
the mock photon count {\sl Chandra} \xray map of the region around the
simulated cold front.  The map is shown in Figure~\ref{chandra_map}.
We assumed a 50~ksec exposure in the $0.5-4$ keV band, cluster
redshift of $z=0.05$, and a flat background level of
$4.3\times10^{-6}$~cnts~s$^{-1}$~pixel$^{-1}$. The figure shows the
full $16^\prime\times 16^\prime$ ACIS-I field of view binned to
$2^{\prime\prime}$ pixels. The color represents photon counts per
pixel in log$_{10}$ scale. The figure shows that the cold front is
detectable by {\sl Chandra} and appears remarkably sharp in the photon
count map.  The boundary of the front spans $\sim 0.5\hMpc$,
comparable to the size of the observed cold front in A3667.  At larger
angles, the sharp discontinuity widens and then gradually disappears,
the onset of Kelvin-Helmholtz instability also seen in A3667; however,
the appearance of the simulated cold front seems to be more irregular
than that of the observed front (see also Fig.~\ref{dm_temp} and $\S$5
for more discussions).

To examine the processes that produce and then destroy this cold
front, we created a movie of the temperature, density, and X-ray
surface brightness evolution similar to the maps in
Figure~\ref{xsbt_map}.  The movie is based on 142 outputs finely
spaced between $z=0.66$ and 0.26.  Examination of the evolution showed
that the large-scale cold front arises when a merging dense subclump
is moving through hotter ambient intergalactic gas of the merging
system.  The hotter temperatures are due to the merger
shocks\footnote{\citet{bialek02} pointed out that the gas sloshed out
  of its potential well can expand and adiabatically cool, which would
  also enhance the temperature difference between clump's and ambient
  gas.}.  The shocks are driven by supersonic motions of the merging
clusters and thus propagate away from them (the situation is similar
to that of a supersonically moving piston in a gas tube). The shocks
appear to originate as cluster pass each other for the first time when
their velocity is greatest.  The clumps thus trail the bow shocks they
produce and the cold gas of clusters is not actually shocked until
much later when the cold front is disrupted. Two additional weaker bow
shocks are produced when the clusters pass each other for the second time.
These secondary shocks play a role in heating the cold gas and
disrupting the front.  As the cold front observed in the temperature
map is located at the large distance from the cluster center the front
exists approximately for a time it takes for its sub-cluster to travel
back to the center and merge plus the time required for the secondary
shock to propagate from the center to the front. This is roughly equal
to the dynamical time of the cluster ($\sim 10^9$~years).
Kelvin-Helmholtz instability and merger-induced turbulence also appear
to contribute to eventual disruption of the front. Based on this
estimate, the frequency of large cold fronts should be somewhat
smaller than the frequency of major mergers in progress because the
cold front may appear only in certain projections (projection
direction perpendicular to the plane of the merger).  Cold fronts thus
appear to be a common dynamic phenomenon accompanying merger events
during cluster formation.

\subsubsection{Small-scale Cold front from Minor merger}
\label{sec:small}

Examining outputs of the simulations, we find that large-scale sharp
gradients of gas temperature most often correspond to the boundaries
between the relatively cold intracluster gas of merging clumps and
hot, already shocked gas surrounding them. This is in agreement with 
interpretation of \citet{markevitch00} and \citet{vikhlinin01}.
We found many smaller cold fronts in frequent 
minor mergers.  For example, a small-scale cold front associated with
the merging subclump to the left of the cluster center can be seen in
the $z=0$ panel of Figure~\ref{xsbt_map}.  This cold front has a size
of $\approx 50h^{-1}$~kpc and temperature of $\lesssim 2$~keV, so this
is essentially a poor group.  Here, unlike the advanced merger stage
discussed in the previous section, the merging subclump is on the
first approach to the cluster center and has not yet suffered major
tidal disruption.  The motion of the dense clump of gas compresses the
ICM upstream of the front.  The compression sharpens and enhances the
amplitude of gas density and temperature gradients across the front.
The surface brightness, temperature and pressure profiles across this
cold front are shown in the right panels of Figure~\ref{xsbt_pro}.
Here again, the simulated profiles reproduce all the main features of
the observed cold fronts.

The smaller fronts would be much more difficult to observe because
they have both much lower \xray luminosity and surface brightness
contrast; however, they can be observed in the nearby clusters.
Recently, \citet{forman02} presented examples of small-scale cold
fronts in the Fornax cluster which have a size of $\approx 20-40$~kpc
and temperatures of $<1$~keV.  These observed properties are very
similar to those found in our simulations.  Note also that the
relatively cooler gas of the merging sub-clump is trailed by
low-entropy ($\sim$1-2~keV) intergalactic gas in the direction of the
subclump's motion (see also the entropy map in Figure~\ref{clevol}).
Such elongated regions of low-temperature gas trailing a subclump, if
observed, can be identified with regions where filamentary material is
accreting onto the cluster.  The filamentary structures visible in the
temperature map, however, does not appear in the \xray surface
brightness map.

\subsection{Dynamics of gas motions}
\label{sec:dynamics}

Detailed understanding of gas dynamics in the vicinity of cold fronts is
important for placing meaningful constraints on the
physics of the ICM, since the power of these constraints relies on the
validity of the underlying assumptions of the dynamical model.

Figure~\ref{dm_temp} presents projected DM density maps with overlayed
contours of emission-weighted temperature (top panels) and
emission-weighted temperature maps with overlayed gas velocity fields
(bottom panels) for the two cold fronts discussed in the previous
section.  The velocity fields were constructed by averaging velocities
in $\approx 78\hkpc$ slice centered on the DM subclump associated with
cold front.  The figure reveals a variety of high-velocity streams
arising during cluster merger.  In the large cold front
($z=0.43$) shown in the left panels, the dominant feature of the
velocity field is the collision of two gas flows on the shock
boundary.  Note that there is no indication of a regular gas flow
around the cold front boundary.  Instead, the cold gas of the front
appears to be moving together with the surrounding hot gas.  Moreover,
it is clear that the subclump is in the process of tidal disruption:
the velocity of its gas is highly non-uniform and the peak of the DM
density is displaced with respect to the center of the front.  In the
$z=0$ minor merger (right panels) the ICM of the main cluster is
considerably more relaxed (typical velocities of just $\sim
100-300\mkms$).  The bulk flow of gas associated with the merging
subclump is also quite regular.  Although in this case the assumption
of regular flow would be closer to truth, the figure shows that the
flow on small-scales around the cold-hot interface is still quite
complicated which may hamper development of Kelvin-Helmholtz
instabilities.

Finally, we would like to note that we performed a similar search for
cold fronts in the SCDM cluster simulation.  We found several cold
fronts with properties similar to those found in the $\Lambda$CDM run.
Here again the features were associated with mergers in different
stages.  In the two clusters that we analyzed, we identified many
``small-scale'' cold fronts many of which are located in the outskirts
of the main cluster progenitor. 

\section{Discussion and conclusions}
\label{sec:discussion}

We used high-resolution $N$-body$+$gasdynamics cluster simulations in
CDM models to search for counterparts of {\em ``cold fronts''} within
cluster ICM recently discovered in {\sl Chandra} observations.  The
observed cold fronts are very sharp features in the X-ray surface
brightness and temperature maps found in several clusters.  These
features are characterized by an increase in surface brightness by a
factor $\gtrsim 2$ accompanied by a {\it drop} in temperature of a
similar magnitude over the scale of 10-50~kpc.  The existence of such
sharp features can put interesting constraints on the properties of
the intracluster medium (ICM), including the efficiency of energy
transport \citep{ettori00} and the magnetic field strength in the ICM
\citep{vikhlinin01}.  In this paper, we investigate whether such features
can be produced naturally in clusters forming in hierarchical models
and study their origin, structural properties and detectability.

We find that cold fronts appear to be common and often arise in major
and minor cluster mergers.  In the preceding sections we discussed two
specific cases of cold fronts found in the simulation of the
$\Lambda$CDM cluster. In the most spectacular case, the cold front
arises in a major merger 
when the merging subclump, moving slightly supersonically, is
undergoing tidal disruption (see distribution of DM and velocity field
in Fig.~\ref{dm_temp}). The motion of the gas, which appears to be in
the process of being ``sloshed out'' of the potential well of its DM
halo, drives the shock and leads to a steepening of density and
temperature gradients which produces the cold front.  The surface
brightness and temperature profiles across the front show that a drop
of temperature over the scale of $20-30\hkpc$ is accompanied by a jump
in surface brightness by a factor of $\gtrsim 2$, behavior very
similar to the profiles of observed cold fronts. In addition, we find 
that similarly to observations pressure changes smoothly accross
cold fronts in simulations. The spatial extent of
the cold front in simulation ($\sim 0.5\hMpc$) is comparable to the
extent of two prominent observed cold fronts in clusters A2142 and
A3667 \citep{markevitch00,vikhlinin01}. Note also that the cold front
is preceded by a bow shock ($\sim 200$~kpc from the cold front). In
this case the jump in surface brightness by a factor of $\gtrsim 5$ is
accompanied by an {\it increase} in temperature of a similar magnitude
over the scale of $\sim 40$~kpc. The shock front has a much lower
\xray surface brightness than the gas around the cold front and
therefore would be difficult to observe. Nevertheless, indications of
a similar bow shock preceding the cold front were found in A3667.

In another discussed case, the cold front arises in a minor merger
when a merging subclump reaches the inner regions of the larger
cluster and gas in front of the subclump is significantly compressed.
The compression sharpens and enhances the amplitude of gas density and
temperature gradients across the front. Here the subcluster is on the
first approach to the main cluster's core and is not yet significantly
disturbed by tides. The relatively cooler gas of the merging sub-clump
is trailed by low-entropy ($\sim$1-2~keV) intergalactic gas in the
direction of the subclump's motion. Such elongated regions of
low-temperature gas trailing a subclump, if observed, can therefore be
identified with regions where filamentary material is accreting onto
cluster and could potentially be targets for studies of properties of
intergalactic material. Unfortunately, the \xray surface brightness of
this gas is very low and detecting such regions will be a challenge.
The best chance of observing such gas in candidate regions would be
X-ray and UV metal absorption lines in spectra of bright background
sources \citep[e.g.,][]{hellsten98}.

Cold fronts in both observed and simulated clusters seem to come in
all shapes and sizes and are rather frequent, as reports of many new
instances of cold fronts are published
\citep[e.g.][]{sun02,forman02,markevitch02}.  In the two clusters that
we analyzed, we identified many ``small-scale'' cold fronts many of
which are located in the outskirts of the main cluster progenitor.
Most of these cold fronts would probably not be detected in distant
clusters with typical exposures and field sizes of Chandra
observations, although they are similar to those observed in the
Fornax cluster \citep{forman02}.  The fact that large-scale ($\sim
0.5$~Mpc) cold fronts arise in major mergers and survive for
approximately dynamical time ($< 1$~Gyr) implies that their frequency
will be similar to the frequency of major mergers. The latter is a
very strong function of redshift ($\propto (1+z)^{\sim 3}$,
\citep[see, e.g.][]{gottlober01}) and may not be very high at $z=0$
(depending on cosmology and the power spectrum normalization), but be
increasingly larger for higher redshift clusters.  Indeed, the
large-scale cold front that we discussed occurs at $z\approx 0.4$.
Observationally, cluster ZW3146 at $z=0.29$ exhibits several cold
front like features \citep{forman02}. More recently,
\citet{markevitch02} reported results of a systematic search for cold
fronts in archival {\sl Chandra} observations of 30 apparently relaxed
clusters. Surprisingly, they find cold-front like features in the
large fraction of the clusters.  Larger samples of both observed and
simulated clusters will be useful to determine the frequency of cold
fronts.

Our results support the interpretation of cold fronts as boundaries
between hot ICM of the main cluster and ``the dense subcluster core
that has survived a merger and ram pressure stripping by the
surrounding shock-heated gas'' \citep{markevitch00}. Based on this
interpretation, \citet{vikhlinin01} proposed the detailed dynamical
model of the cold front observed in cluster A3667 assuming a spherical
dense gas subclump in a spherical dark matter potential moving through
the uniform ICM. This model was used to place constraints on the
magnetic field configuration and strength and plasma conductivity in
the vicinity of cold front. The conductivity can be constrained simply
by the sharpness of the observed front if a certain lifetime is
assumed for the front. The constraint on the magnetic field
configuration, on the other hand, is obtained from the fact that no
indication of Kelvin-Helmholtz instabilities is present in
observations.  Such instabilities are expected to develop quickly for
a regular high-velocity flow of hot surrounding gas around the dense
subclump along the cold front boundary. The power of these constraints
relies on the validity of the underlying assumptions. Of these, the
assumption of regular flow is particularly worrisome, given the
complicated flow structure apparent from entropy and temperature maps
shown in Figs.~\ref{clevol} and~\ref{xsbt_map}.

Indeed, Figure~\ref{dm_temp} shows that this assumption fails in the
case of cold fronts identified in our cluster simulations.  Our
simulations show that dynamics of gas around the simulated cold front
is much more complicated than that assumed in Vikhlinin et al.'s
model.  In particular, the flow of gas around the front is not laminar
and in general is not parallel to the front.  Vikhlinin et al., on the
other hand, assumed a laminar flow similar to flow of gas around a
blunt body.  In this case the gas flow lines are parallel to the front
- the conditions ideal for development of the Kelvin-Helmholtz
instabilities.  The velocity field shown in Fig.~\ref{dm_temp} is very
different and the conditions are probably not as conducive to the
Kelvin-Helmholtz instability as those assumed in Vikhlinin et al.'s
model.  On the other hand, the cold front in A3667 that they model has
a much more regular appearance than the simulated cold front shown in
Fig.~\ref{chandra_map}.  But, such detailed comparison is difficult
because processes discussed by Vikhlinin et al. would be poorly
resolved in our simulations, which cannot resolve well perturbations
smaller than 20~kpc.  It is also difficult to draw conclusions about
how generic the differences are (i.e., whether the flow pattern
assumed in Vikhlinin et al. can arise in simulated clusters).  These
questions will be addressed in future studies using considerably
higher resolution simulations and a large sample of simulated
clusters, respectively.

To conclude, our results indicate that cold fronts are non-equilibrium
transient phenomena arising in cluster mergers. They occur when the
relatively cold gas of the merging subclump is moving fast with
respect to the hotter gas of the main cluster.  Both simulated cold
fronts discussed in the paper would be detectable in a {\sl Chandra}
observation of $\sim 50$~ksec exposure, if the clusters were located
at a moderately low redshift (e.g., $z\lesssim 0.1$, see
Fig.~\ref{chandra_map}). Cluster mergers (especially minor) are common
and the fact that we were able to identify several cold fronts in the
simulations of only two clusters means that we can expect many more
cold fronts to be discovered in the near future
\citep[e.g.,][]{sun02}. If the origin and properties of these features
are understood, they can provide powerful constraints on the frequency
of cluster mergers and properties of the intracluster medium. The
analysis presented here is a first step in studying the origin and
properties of cold fronts. The presented results should be useful in
assessing the validity of the assumptions in the detailed modeling of
cold fronts and developing more sophisticated models.

\acknowledgements 
We would like to thank Stefan Gottl\"ober for providing the initial
conditions for the $\Lambda$CDM cluster simulation and computing
resources used for performing the analysis of the simulations. We
appreciate useful discussions with Gus Evrard, Alexey Vikhlinin, Maxim
Markevitch, Joe Mohr, and Anatoly Klypin.  We also thank the anonymous
referee for constructive comments on the manuscript.  The work
presented here was partially supported by NASA through a Hubble
Fellowship grant from the Space Telescope Science Institute, which is
operated by the Association of Universities for Research in Astronomy,
Inc., under NASA contract NAS5-26555 and by NSF through funding of the
Center for Cosmological Physics at the University of Chicago (NSF
PHY-0114422).  DN thanks John Carlstrom for his support through NASA
LTSA grant NAG5--7986.


\end{document}